\documentstyle[12pt]{article} \textwidth 14.5cm \textheight 
20cm
\begin{document} 
\newcommand{\add}{\addtocounter{eqncnt}{1}}
\newcounter{eqncnt}[section]
\newcommand{\al}{\mbox{$\alpha$}}
\newcommand{\med}[1]{g_{#1}} 
\newcommand{\meu}[1]{g^{#1}}
\newcommand{\be}{\begin{equation}} 
\newcommand{\ee}{\end{equation}\add}
\newcommand{\bea}{\begin{eqnarray}}

\renewcommand{\theequation}{\arabic{section}.\arabic{equation}}

\newcommand{\eea}{\end{eqnarray}\add}
\newcommand{\und}{\underline}
\newcommand{\D}{\displaystyle}
\newcommand{\hs}{\hspace*{2in}}
\newcommand{\hqq}{\hfill\qquad}
\newcommand{\noin}{\noindent}

\begin{center}
{\sc AN ASYMPTOTIC ANALYSIS OF SPHERICALLY SYMMETRIC PERFECT FLUID 
SIMILARITY SOLUTIONS}
\end{center}

\vskip .25in
 
\begin{center}
B. J. Carr\\
Astronomy Unit, Queen Mary \& Westfield College\\
 University of London, London E1 4NS, UK\\
Yukawa Institute for Theoretical Physics, Kyoto University, Kyoto 606, Japan\\
and\\
A. A. Coley\\
Department of Mathematics and Statistics, Dalhousie University\\
Halifax, Nova Scotia  B3H 3J5  Canada
\end{center}

\vskip .5in 
 
\begin{abstract}
The asymptotic properties of self-similar spherically symmetric
perfect fluid solutions with equation of state $p=\alpha \mu$ ($-1<\alpha<1$) are described.  We prove that for large
and small values of the similarity variable, $z=r/t$, all such solutions must have
an asymptotic power-law form. They are associated either with an 
{\it exact} power-law solution, in which case the $\alpha >0$ ones are asymptotically Friedmann, 
asymptotically Kantowski-Sachs or asymptotically static, or with an {\it approximate} power-law solution, in which
case they are asymptotically quasi-static for $\alpha >0$
or asymptotically Minkowski for $\alpha >1/5$. We also show
that there are solutions whose asymptotic behaviour is associated with {\it finite}
values of $z$ and which depend upon powers of $ln z$. These correspond either to a second family of asymptotic
Minkowski solutions for $\alpha>1/5$ or to solutions that are asymptotic to a central singularity for $\alpha>0$ .
The asymptotic form of the solutions is given in all cases, together with the number of
associated parameters. This forms the basis for a complete classification of all $\alpha >0$ self-similar solutions.
There are some other asymptotic power-law solutions associated with negative $\alpha$, but the physical significance of
these is unclear. 
\end{abstract}

\setcounter{equation}{0}
\section{Introduction}
Spherically symmetric solutions 
to Einstein's equations admitting a self-similarity of the first kind
are characterized by the fact that the spacetime possesses a homothetic Killing 
vector. This means that they can be 
put into a form in which every dimensionless variable is a 
function of some dimensionless combination of the cosmic time 
coordinate $t$ and the comoving radial coordinate $r$. Such solutions have attracted considerable attention in recent decades.
This is partly for mathematical reasons, since the governing equations reduce to ordinary
differential equations from partial differential equations.
In particular, in the case of a perfect fluid, the only barotropic equation of  state compatible with the similarity
assumption is one of  the form $p = \alpha \mu$, where $\alpha$ is a constant,
and the equations then simplify still further. However, it is also for physical reasons, since such solutions 
play a crucial role in many cosmological and astrophysical contexts and are important 
in the context of the ``critical" phenomena discovered in recent
gravitational collapse calculations (Evans and Coleman 
1994, Carr et al. 1999, Neilsen and
Choptuik 1998).
Indeed the ``similarity hypothesis" proposes that spherically symmetric solutions may naturally evolve 
to similarity form in a wide variety of situations. The arguments for this have been reviewed by  
Carr \& Coley (1999a). 

The possibility that self-similar models may be singled out in this
way from more general spherically symmetric solutions means that 
it is essential to study the full family of such solutions. 
Two main approaches have been used in such studies.
The first uses the ``comoving'' approach, 
pioneered by Cahill \& Taub (1971), in which the
coordinates are adapted to the fluid 4-velocity vector and 
the solutions are  parametrized 
by the
similarity variable $z=r/t$. At a given value of $t$, this
specifies the spatial profile of the various quantities. 
At a given
value of $r$ (i.e. for a given fluid element), it 
specifies their time
evolution. The second uses the ``homothetic'' approach, 
introduced by Bogoyavlensky (1977). In this, the 
coordinates are adapted to
the homothetic vector and solutions can be treated as 
orbits in
a compactified 3-dimensional phase space.

The purpose
of this paper is to delineate all possible asymptotic behaviours of spherically symmetric $p=\alpha \mu$
similarity solutions using the comoving approach. This provides the first step in our ``complete" classification of all such
solutions with $\alpha>0$ in another paper (Carr \& Coley 1999b; CC). The classification is complete, subject to the
requirement that the solutions are physical everywhere and do not contain shocks. However, our claim for
completeness in that paper is based on the assumption that all solutions must have an asymptotic form in which any
dimensionless quantity has a power-law dependence on $z$ at large and small values of $z$ or a power-law
dependence on $ln z$ at finite $z$ (whenever this corresponds to zero or infinite
distance from the origin). We do not {\it prove} this assumption in CC but take it as our starting point. We then find that 
the behaviour of solutions in the asymptotic limits can only take one of a few
simple forms for $\alpha>0$: asymptotically Friedmann, asymptotically Kantowski-Sachs, 
asymptotically Minkowski (for $\alpha > 1/5$), asymptotically singular or what
we term asymptotically ``quasi-static".

In this paper we also obtain the asymptotic behaviours but in a different way and, in the process, we provide a rigorous 
{\em proof} of the ``power-law" assumption in the physically important positive $\alpha$ case. This paper therefore
crucially complements the classification paper. Indeed we sometimes refer to equations in CC below. Both
papers end up with the same asymptotic behaviours for
$\alpha>0$ but the analysis in this paper is more rigorous and more comprehensive. In addition, this paper also covers the negative 
$\alpha$ case. However, we do not consider the physical significance
of any of the solutions in this paper. The  positive $\alpha$ solutions are discussed in more detail elsewhere (CC, Carr et al. 1999).
The physical significance of the negative $\alpha$ solutions are less clear; some applications are discussed in Carr \& Coley (1999a)
and we note that for $-1<\alpha<-1/3$ the asymptotically Friedmann (AF) solutions (for example -- see Table 1) are
inflationary (Olive, 1990)

It should be stressed that Goliath et al. 
(1998a \& b; GNU) have also implicitly provided a 
classification of spherically symmetric self-similar solutions with positive $\alpha$
using the homothetic approach. However, 
the nature of their classification is rather different in that they 
emphasize the equilibrium points of
the associated dynamical system. These always lie on the 
{\it boundary} of the 3-dimensional phase space, whereas the 
condition
$z\rightarrow\infty$ emphasized by CC usually corresponds to the {\it interior} of the phase
space and so does not play a crucial role in the GNU analysis
(for example, there are no equilibrium points corresponding to CC's
asymptotically quasi-static solutions which occur as $|z|\rightarrow\infty$). However, some of these equilibrium points do have a
straightforward connection to CC's analysis. Thus what GNU term the ``F point"
corresponds to solutions which are asymptotically Friedmann as $z\rightarrow \infty$, their ``$C^0$ point" corresponds to 
solutions which are asymptotically Friedmann as $z\rightarrow 0$ (or regular at the origin), their ``T point"
corresponds to the static solution, and their ``K point"
corresponds to (asymptotically Kasner) solutions with a central singularity
(occuring at a finite value  of $z$). 
The two approaches are therefore complementary and the value of 
using a combined approach has been emphasized by Carr et al. (1999).

The outline of this paper is the following. In Section 2 we
introduce the general spherically symmetric similarity 
equations for a perfect fluid and highlight the features necessary for 
our analysis. In Section 3 we provide the 
analysis itself; this comprises an exhaustive case-by-case consideration of the different possible behaviours
of various key functions. In Section 4
we draw some general conclusions and point out some more subtle issues which are not covered explictly in Section 3.

\setcounter{equation}{0}
\section{Spherically Symmetric Similarity Solutions}

In the spherically symmetric situation one can introduce a time coordinate
$t$ such that surfaces of constant $t$ are orthogonal to fluid flow lines
and comoving coordinates ($r,\theta ,\phi$) which are constant along each
flow line. The metric can then be written in the form
\be
\label{lelement}
        ds^{2}=e^{2\nu}\,dt^{2}-e^{2\lambda}\,dr^{2}-R^{2}\,d\Omega^{2},\;\;\;
        d\Omega ^2 \equiv d\theta^{2}+\sin^{2}\theta \,d\phi^{2} 
\ee
where $\nu$, $\lambda$ and $R$ are functions of $r$ and $t$. For a perfect 
fluid the Einstein equations are
\be
        G^{\mu \nu}=8\pi[(\mu+p)U^{\mu}U^{\nu}-p\,g^{\mu \nu}]
\ee
where $\mu(r,t)$ is the energy density, $p(r,t)$ the pressure, 
$U^{\mu}=(e^{-\nu},0,0,0)$ is the comoving fluid 4-velocity, and we choose
units in which $c=G=1$. The equations have a first integral
\be
\label{firstint}
   m(r,t)=\mbox{$\frac{1}{2}$} R\left[ 1+e^{-2\nu}
   \left(\frac{\partial R}{\partial t}
   \right)^{2} - e^{-2\lambda}\left(\frac{\partial R}
   {\partial r}\right)^{2}\right]
\ee
and this can be interpreted as the mass within comoving radius $r$ at 
time $t$: 
\be
\label{massfunct}
   m(r,t)=4\pi\int_{0}^{r}\mu R^{2}\frac{\partial R}{\partial r'}\,dr'.
\ee
Unless $p=0$, this quantity decreases with increasing $t$ because of the
work done by the pressure.

A  self-similar solution is one in 
which the spacetime admits a homothetic Killing vector \mbox{\boldmath$\xi$} such that
\be
        \xi_{\mu ;\nu}+\xi_{\nu ;\mu} =2g_{\mu \nu}.
\ee 
This means that the solution is unchanged by a transformation of the form
$t\rightarrow at$, $r\rightarrow ar$ for any constant $a$. 
Solutions of this type were first
investigated by Cahill \& Taub (1971), who showed that by a suitable
coordinate transformation they can be put into a form in which all
dimensionless quantities such as $\nu$, $\lambda$,
\be
   S\equiv\frac{R}{r},\;\;\;\;M\equiv\frac{m}{R},\;\;\;\;
   P\equiv pR^{2},\;\;\;\;W\equiv\mu R^{2}
\ee
are functions only of the dimensionless variable $z\equiv r/t$. Then we have
\be
   \frac{\partial\;}{\partial t}=-\frac{z^{2}}{r}\frac{d\;}{dz},\;\;
   \;\; \frac{\partial\;}{\partial r}=\frac{z}{r}\frac{d\;}{dz},
\ee
so the field equations reduce to a set of ordinary differential
equations in $z$. Another important quantity is the function
\be
\label{velocity}
        V(z)=e^{\lambda -\nu}z,
\ee
which represents the velocity of the surfaces of 
constant $z$ relative to the fluid. These surfaces have the equation $r=z t$ and therefore
represent a family of spheres moving through the fluid. The spheres contract relative to the fluid for $z<0$ and expand 
for $z>0$. Special significance is attached to values of $z$ for which
$|V|=1$ and $M=1/2$. 
The first corresponds to a Cauchy horizon (either a black hole 
event horizon
or a cosmological particle horizon), 
the second to a black hole or cosmological apparent horizon.

The only barotropic equation of state compatible with the similarity ansatz is
one of the form $p=\alpha \mu$ ($-1\leq\alpha\leq 1$). It is convenient to 
introduce a dimensionless function $x(z)$ defined by
\be
\label{xdef}
        x(z)\equiv (4\pi \mu r^{2})^{-\alpha/(1+\alpha)}.
\ee
The conservation equations $T^{\mu \nu}_{\;\;\;\; ;\nu}=0$ can then be 
integrated to give
\be
\label{metrictt}
        e^{\nu}=\beta x z^{2\alpha/(1+\alpha)}
\ee
\be
\label{metricrr}
        e^{-\lambda}=\gamma x^{-1/\alpha} S^{2}
\ee
where $\beta$ and $\gamma$ are integration constants. The remaining field 
equations reduce to a set of ordinary differential equations in $x$ and 
$S$:
\be
   \ddot{S}+\dot{S}+\left(\frac{2}{1+\alpha}\frac{\dot{S}}{S}
   -\frac{1}{\alpha}\frac{\dot{x}}{x}\right)
   [S+(1+\alpha)\dot{S}]=0,
\ee
\be
   \left(\frac{2\alpha\gamma^{2}}{1+\alpha}\right)S^{4}
   +\frac{2}{\beta^{2}}\frac{\dot{S}}{S}\,x^{(2-2\alpha)/\alpha}
   z^{(2-2\alpha)/(1+\alpha)} - \gamma^{2}S^{4}\,\frac{\dot{x}}{x}
   \left(\frac{V^{2}}{\alpha}-1\right) = (1+\alpha)x^{(1-\alpha)/\alpha},
\ee
\be
   M=S^{2}x^{-(1+\alpha)/\alpha}\left[1+(1+\alpha)\frac{\dot{S}}{S}
   \right],
\ee
\be
   M=\frac{1}{2}+\frac{1}{2\beta^{2}}x^{-2}z^{2(1-\alpha)/(1+\alpha)}
   \dot{S}^{2}-\frac{1}{2}\gamma^{2}x^{-(2/\alpha)}S^{6}
   \left(1+\frac{\dot{S}}{S}\right)^{2},
\ee
where the velocity function is given by
\be
   V=(\beta\gamma)^{-1}x^{(1-\alpha)/\alpha}S^{-2}
   z^{(1-\alpha)/(1+\alpha)}
\ee
and an overdot denotes $zd/dz$.

We can 
envisage how these equations generate solutions by working in the
$3$-dimensional $(x, S, \dot{S})$ space (Carr and Yahil 1990). At any point
in this space, for a fixed value of $\alpha$, eqns (2.14) and (2.15) give the value of z; eqn (2.13) then gives the 
value of $\dot{x}$ unless $|V|=\sqrt{\alpha}$ and eqn (2.12) gives the value of $\ddot{S}$. Thus the equations generate a
vector field $(\dot{x}, \dot{S}, \ddot{S})$ and this specifies an integral curve at each point of the 3-dimensional space.
Each curve is parametrized by $z$ and represents one particular similarity solution.
This shows that, for a given equation of state parameter $\alpha$, there is a
$2$-parameter family of spherically symmetric similarity solutions. Special significance
is attached to points where $|V| = \sqrt{\alpha}$, which specifies a 2-dimensional 
surface in $(x, S, \dot{S})$ space. This corresponds to a sonic point, so there can be a discontinuity in the pressure 
gradient (i.e. the value of $\dot{x}$) here. The requirement that solutions be regular at the sonic point (in the sense that
they can be extended  beyond there) places severe restrictions on the nature of the solutions (Bogoyavlensky 1977, 
Bicknell \& Henriksen 1978, Carr \& Yahil 1990, Ori \& Piran 1990) but these
restrictions are not relevant to the present asymptotic analysis.

\setcounter{equation}{0}
\section{Asymptotic Classification}

In this section we will provide a complete analysis of spherically symmetric
similarity solutions in the limits $R \rightarrow 0$ and $R \rightarrow \infty$. (We will assume 
$z>0$ throughout since the $z<0$ solutions are just the time reverse of the $z>0$ ones.) This may
correspond to $z\rightarrow 0$, $z\rightarrow \infty$ or $z\rightarrow z_*$ (finite)
since the correspondence between the comoving radial coordinate $r$ and the physical distance
$R$ may not be straightforward. In particular, $r\rightarrow0$ need not imply $R\rightarrow0$ (eg. some dust 
solutions), $R\rightarrow0$ need not imply $r\rightarrow0$ (eg. black hole singularities), $r \rightarrow \infty$ need not
imply 
$R \rightarrow \infty$ (eg. the Kantowski-Sachs solution) and $R \rightarrow \infty$ need not imply 
$r \rightarrow \infty$ (eg.
asymptotically Minkowski solutions). We categorize solutions according to whether
$V\rightarrow 0$, $V\rightarrow \infty$ or $V\rightarrow V_*$ (finite) in each of 
the $z$ limits. A key role is played by eqn (2.13), which can be written as
\be
\frac{\dot{x}}{x} = \frac{2V^2\alpha}{V^2-\alpha}\frac{\dot{S}}{S} + \frac{2\alpha^2}{1+\alpha}\frac{1}{V^2-\alpha} - \xi,
\;\;\; \xi \equiv \frac{\alpha (1+\alpha)}{\gamma^2} \frac{S^{-4}x^{(1-\alpha)/\alpha}}{V^2-\alpha} . 
\ee
One has different possible behaviours according to whether $\xi \rightarrow 0$, $\xi \rightarrow \infty$ or $\xi \rightarrow
\xi_*$ (finite). In principle we need to analyse 27 cases,
corresponding to the three limiting behaviours of each of $z$, $V$ and $\xi$. However, in practice, 
many of these do not lead to a consistent solution. In the discussion below, we usually only derive the 
self-consistent ones. 

Our procedure is as follows: for each limiting value of {$z$, $V$ and $\xi$}, we use eqn (3.1) to eliminate
$\dot{x}/x$ in eqn (2.12) and thereby obtain a second order ODE for $S$. Not every solution of
this ODE may be consistent with the other equations, so we first confirm that $\xi$ has
the correct limiting value and then check that the integral condition associated with eqns
(2.14, 2.15) is satisfied. This procedure not only identifies the possible asymptotic
behaviours but also indicates the number of free parameters in each case.  The results are summarized in Table 1.

\vskip .3in

\begin {tabular}{|c|c|c|c|} \hline
& $z\rightarrow 0$ & $z\rightarrow z_*$ & $z \rightarrow \infty$ \\ \hline
& AKS (0/1) ($1>\alpha>-1/3$) & & \\
$V\rightarrow 0$ & AF (1) ($1>\alpha>-1/3$) & No soln & AF (1) ($-1<\alpha<-1/3$)   \\ 
& ES (0) ($1>\alpha>0$) & & AKS (1) ($-1<\alpha<-1/3$) \\ 
& AX (1) ($-0.17<\alpha <-1/7$) & & AX (1) ($1>\alpha >-1/7$)\\ \hline
$V\rightarrow V_*$ & No soln & AM (2) ($1>\alpha>1/5$) & AM (1) ($1>\alpha>1/5$) \\ 
& & AKS (1) ($1>\alpha>0$) & AM (1) ($-1<\alpha<0$) \\ \hline
& & & AQS (2) ($1>\alpha>0$) \\
$V\rightarrow \infty$ & AKS (1) ($-1<\alpha<-1/3$) & AK (2) ($1>\alpha>-1/3$) & AKS (0/1) ($1>\alpha>-1/3$) \\
& AF (1) ($-1<\alpha<-1/3$) & & AF (1) ($1>\alpha>-1/3$) \\
& & & AY (1) ($-1<\alpha<0$) \\ \hline
\end {tabular}

\vskip .25in

\noin
Table 1. {\it This summarizes the asymptotic (A) behaviour of spherically symmetric perfect fluid similarity 
solutions for different ranges of $\alpha$.  The number of arbitrary parameters for each solution 
is given in parentheses. KS, F, M, K, ES and QS
denote Kantowski-Sachs, Friedmann, Minkowski, Kasner, Exact Static and Quasi-Static respectively.
X and Y correspond to the two new solutions in which either $V$ or $M$ is negative. The AKS solution is
described by one parameter for $\alpha>0$ and $-1<\alpha<-1/3 $ but there is only the exact KS solution for $-1/3<\alpha<0$;
$V$ and $M$ are positive only for $-1<\alpha<-1/3$.}

\vskip .3in

In the following analysis a differential equation of the form

\be
\ddot{S} + a\dot{S}+ b\frac{\dot{S}^2}{S} =0,
\ee
where $a$ and $b$ are constants, will frequently arise. When $a \ne 0$, we can divide through by $S$ and
integrate this equation to obtain

\be
\dot{S}S^b \sim z^{-a}.
\ee
If $b \ne -1$, we can then integrate a second time
to obtain the exact solution

\be
S = (c_1 + c_2 z^{-a})^{1/(1+b)},
\ee
where $c_1$ and $c_2$ are integration constants. 
The dominant asymptotic solution is then determined by the sign of $a$.
Note that the first order perturbation to the dominant solution (which is not derived in the present analysis) 
will often be larger than the subdominant solution.

\vskip .3in

$\bullet$ $V\rightarrow 0$ as $z \rightarrow 0$

In this case, eqn (3.1) becomes
\be
\frac{\dot{x}}{x} = - \frac{2\alpha}{1+\alpha}- \xi,  \;\;\;  \xi \equiv -\frac{1+\alpha}{\gamma^2}
S^{-4}x^{(1-\alpha)/\alpha}. 
\ee
If $\xi \rightarrow 0$, this implies $x \sim z^{-2\alpha/(1+\alpha)}$, so eqn (2.12) gives
\be
\ddot{S} + \left(\frac{5+3\alpha}{1+\alpha}\right) \dot{S}+ \frac{2\dot{S}^2}{S}+ \frac{2S}{1+\alpha} =0.
\ee
Defining $\sigma(z) \equiv zS(z)$, this becomes

\be
\ddot{\sigma} - \left(\frac{1+3\alpha}{1+\alpha}\right) \dot{\sigma}+ \frac{2\dot{\sigma}^2}{\sigma} =0.
\ee
This differential equation is of the form of (3.2) and, from eqn (3.4), we therefore obtain the
exact solution

\be
S = z^{-1}[c_1 + c_2 z^{(1+3\alpha)/(1+\alpha)}]^{1/3}.
\ee

For $c_1 = 0$, we obtain the exact power-law solution 
\be
S = Az^{-2/3(1+\alpha)},\;\;\;x=Bz^{-2\alpha/(1+\alpha)},
\ee
where the constants $A$ and $B$ are constrained by eqns (2.14, 2.15) and can be 
taken to be
$1$ with a suitable choice of the scaling constants $\beta$ and $\gamma$ in eqns (2.10, 2.11) [see CC eqn (3.10)].
This corresponds to the k=0 Friedmann model. 
Note that
$V\rightarrow 0$ and $\xi\rightarrow 0$ (as assumed) providing
$\alpha>-1/3$. As indicated in Table (1) and discussed in CC, there is also a 1-parameter family
of solutions asymptotic to this. This involves modifying the constants $A$ and $B$, as well as adding higher
order terms in $z$. However,
demonstrating this depends on a higher order analysis than  presented here. 

For $c_2 = 0$, one
obtains another power-law solution
\be
S = Az^{-1},\;\;\;x=Bz^{-2\alpha/(1+\alpha)},
\ee
which corresponds to a
Kantowski-Sachs (KS) model. Eqns (2.14, 2.15) determine the constants $A$ and $B$ in terms of
$\alpha$ once $\beta$ and $\gamma$ are specified [see CC eqn (3.18)]. However, as explained in CC, one requires 
$\beta$ and $\gamma$ to be imaginary for $\alpha>0$, which means that $V$ is negative (i.e. the 
solution is tachyonic).
Furthermore, eqn (2.14) implies that $M$ is negative for $\alpha>-1/3$, so the solution given by
 eqn (3.10) may be unphysical.
Nevertheless, $V\rightarrow 0$ and $\xi\rightarrow 0$ providing
$\alpha>-1/3$, so this solution is still formally valid. As indicated in Table (1) and discussed
 in CC, one can 
perturb this solution to obtain a
1-parameter family of solutions which asymptote towards it. This involves modifying the constants
$A$ and $B$, as well as adding higher order terms. If both constants $c_1$ and
$c_2$ in eqn (3.8) are non-zero, the $c_1$ term in eqn (3.8) dominates as $z \rightarrow 0$ but 
the other term should not be
interpreted as giving the perturbation to KS.

If $\xi \rightarrow \xi_*$ (constant), then 
 $x \sim S^{4\alpha/(1-\alpha)}$ and eqn (2.12) becomes
\be
\ddot{S} - \left(\frac{1+6\alpha + \alpha^2}{1-\alpha^2}\right) \dot{S} - 2\left(\frac{1+3\alpha}{1-\alpha}\right) 
\frac{\dot{S}^2}{S}=0.
\ee
This is again of the form (3.2) and hence has the exact solution
\be
S = [c_1 + c_2 z^{(1+6\alpha + \alpha^2)/(1-\alpha^2)}]^{(\alpha-1)/(1+7\alpha)}.
\ee
If $c_2 = 0$, this leads to the exact static solution with (cf. Misner \& Zapolsky 1964)
\be
x=x_0(\alpha),\;\;\; S=S_0(\alpha).
\ee
This solution is specified uniquely for a given equation 
of state since eqn (3.5) and eqns (2.14, 2.15) give two independent 
relationships between $x_o$ and $S_o$. [See CC eqn (3.30) for the explicit form of eqn (3.13).] However, these values are real
and the mass is positive only for $\alpha>0$.

If $c_1 = 0$, then
\be
S = A z^{-(1+6\alpha+\alpha^2)/(1+\alpha)(1+7\alpha)},\;\;\;x = B z^{-4\alpha (1+6\alpha+\alpha^2)/(1-\alpha^2)(1+7\alpha)},
\ee
where the constants $A$ and $B$ are related by eqns (2.14, 2.15) and by eqn (3.5). From eqns (2.14) and (2.16) one also has
\be
V \sim z^{-(3\alpha+1)^2/(1+\alpha)(1+7\alpha)},\;\;\;M \sim z^{2(3\alpha+1)(1+6\alpha + \alpha^2)/(1-\alpha^2)(1+7\alpha)},
\ee
so this is only
consistent with the assumption that $V \rightarrow 0$ as $z \rightarrow 0$ if $-1<\alpha<-1/7$.
However, eqns (3.5) and (3.14) require 
$-2\alpha(1+3\alpha)^2/(1-\alpha^2)(1+7\alpha) =
(1+\alpha)\gamma^{-2}A^{-4}B^{-(1-\alpha)/\alpha}$,
and this is only satisfied for $-1<\alpha<-1/7$ if $\gamma$ (and hence $\beta$) are imaginary. In this case, 
$M$ is positive but $V$ is negative (as in the KS case), which might be regarded as unphysical. 
However, one can show that eqns
(2.14, 2,15) are still satisfied, and indeed give the same relationship between $A$ and $B$ as above, providing 
$-1/7>\alpha> 2\sqrt{2}-3 = -0.17$. Since this does not seem to correspond to any well-known solution, we
describe it as asymptotically ``X" in Table (1). If both constants $c_1$ and $c_2$ in eqn (3.12) are
non-zero, then the $c_1$ term dominates as $z \rightarrow 0$ for $\alpha >2\sqrt{2}-3$. Although one 
might regard this as an
asymptotically static solution, it is not equivalent to a perturbed static solution. 

If $\xi \rightarrow \infty$, it can be shown that there is no 
consistent solution.

\vskip .3in
$\bullet$ $V\rightarrow \infty$ as $z \rightarrow \infty$

In this case, eqn (3.1) implies
\be
\frac{\dot{x}}{x} = 2\alpha \frac{\dot{S}}{S}+ \left\{\frac{2\alpha^2}
{1+\alpha}\frac{1}{V^2}\right\} -  \xi,  \;\;\;  \xi \equiv 
\alpha(1+\alpha)\beta^2 x^{(\alpha-1)/\alpha}z^{-2(1-\alpha)/(1+\alpha)}, 
\ee
where we have used eqn (2.16) to eliminate $V$ in the expression for $\xi$ The term in curly brackets is included 
to allow for the possibility that 
$\dot{x}/x 
\rightarrow 0$ and $\dot{S}/S \rightarrow 0$. If $\xi \rightarrow 0$, 
this implies $x \sim S^{2\alpha}$ and so eqn (2.12) becomes
\be
\ddot{S} + \left(\frac{1-\alpha}{1+\alpha}\right)\dot{S} -2\alpha \frac{\dot{S}^2}{S}=0.
\ee
Eqn (3.2) implies that the solution of this equation is given by
\be
S = [c_1 + c_2 z^{(\alpha-1)/(1+\alpha)}]^{1/(1-2\alpha)}.
\ee
The case $c_2 = 0$ gives the 
the exact static model, in which case 
$x_o$ and $S_o$ have the values indicated by eqn (3.13) and one needs the term in curly brackets in eqn (3.16).  
The case $c_1 = 0$ gives another 1-parameter solution, 
\be
S = A z^{(\alpha-1)/(1+\alpha)(1-2\alpha)},\;\;\;x = B z^{2\alpha (\alpha-1)/(1+\alpha)(1-2\alpha)},
\ee
where $A$ and $B$ are related by eqns (2.14, 2.15). However, this leads to
\be
V \sim z^{(1-\alpha)/(1+\alpha)(1-2\alpha)},\;\;\;\xi \sim z^{2\alpha (1-\alpha)/(1+\alpha)(1-2\alpha)},
\ee
so  $V \rightarrow \infty$ and $\xi \rightarrow 0$ as
$z\rightarrow\infty$ only for $\alpha<0$.
In this case, eqns (2.14, 2.15) can be satisfied only 
if $\beta$ and 
$\gamma$ are imaginary, so that $V$ is negative, but $M$ is still positive.
Since this does not seem to correspond to any well-known solution, we
describe it as asymptotically ``Y" in Table (1). 

For $c_1 \ne 0$ and $c_2\ne 0$,
eqn (3.18) leads to a solution
\be
S = A + Bz^{-(1-\alpha)/(1+\alpha)},\;\;\; x = C + D z^{-(1-\alpha)/(1+\alpha)}.
\ee
In this case, the term in curly brackets in eqn (3.16) can still be neglected since 
$\dot{x}/x$ and $\dot{S}/S$ go to zero more slowly than $V^{-2}$. There are two relationships between the four integration 
constants {A, B, C,
D}, eqn (3.21) requiring $D=2\alpha B$ and eqns (2.14, 2.15) giving another relationship, 
so these solutions are described by two independent parameters. Note that $S$ tends to a constant as
$z\rightarrow \infty$  but this constant is different from the value
$S_0$ given by eqn (3.13), so these solutions are described as asymptotically ``quasi-static". Indeed eqns (3.21)
corresponds to the full family of such solutions discussed by CC and also (implicitly) by Ori \& Piran (1990)
and Foglizzo \& Henriksen (1992).

If $\xi \rightarrow \xi_*$ (constant), 
then  $x \sim z^{-2\alpha/(1+\alpha)}$ and so eqn (3.18) again applies. For $\alpha >-1/3$, this leads to either the
Friedmann solution, given by eqn (3.9), or the KS solution, given by eqn (3.10). 
Although the general solution with 
$c_1 \neq 0$ and $c_2 \neq 0$ can be written as
\be
S = Az^{-2/3(1+\alpha)}[1+ Bz^{-(1+3\alpha)/(1+\alpha)}]
\ee
and might therefore be regarded as being asymptotic to the Friedmann solution, the second term is not 
equivalent to the first order perturbation of the Friedmann solution. CC show that this pertubation goes like
$z^{-2(1+3\alpha)/3(1+\alpha)}$ and this dominates the second term in eqn (3.22) as $z\rightarrow \infty$. 
 
If $\xi \rightarrow \infty$, there is no consistent solution.

\vskip .3in
$\bullet$ $V\rightarrow \infty$ as $z \rightarrow 0$

In this case, eqn (3.16) applies. If $\xi \rightarrow 0$, one again obtains eqn (3.18). However, the static solution 
with $c_2=0$ has $V\rightarrow 0$ for all $\alpha$, while eqn (3.20) implies that no solution with  $c_1=0$ has both
$V\rightarrow\infty$ and $\xi \rightarrow 0$. If $\xi \rightarrow \xi_*$, one gets both a Friedmann solution of the form 
(3.9) and a KS solution of the form (3.10) with $V\rightarrow\infty$ and $\xi \rightarrow 0$ providing
$-1<\alpha<-1/3$. Furthermore, unlike the $\alpha >0$ case, the KS solution is now physical in that $M$ and $V$ are
positive. If $\xi \rightarrow
\infty$, there is no consistent solution. 
   
\vskip .3in
$\bullet$ $V\rightarrow 0$ as $z \rightarrow \infty$

In this case, eqn (3.5) applies. If $\xi \rightarrow 0$, one again obtains eqn (3.8) and one gets both a Friedmann
solution and a (physical) KS solution with $V\rightarrow 0$ and $\xi \rightarrow 0$
providing
$-1<\alpha<-1/3$. If $\xi \rightarrow \xi_*$, one gets a solution of the form (3.12). However, the static solution 
with $c_2=0$ has $V\rightarrow \infty$ for all $\alpha$, while eqn (3.5) cannot be satisfied for 
$c_1=0$ because 
\be
\frac{\dot{x}}{x} + \frac{2\alpha}{1+\alpha} = -\frac{2\alpha(1+3\alpha)^2}{(1-\alpha^2)(1+7\alpha)}
\ee
and therefore has the same sign as $\xi$. If
$\xi
\rightarrow
\infty$, there is no consistent solution.    

\vskip .3in
$\bullet$ $V\rightarrow \infty$ as $z \rightarrow z_*$

In this case, eqn (3.1) implies
\be
\frac{\dot{x}}{x} = 2\alpha \frac{\dot{S}}{S} -  \xi,  \;\;\;  \xi \equiv \alpha(1+\alpha)\beta^2 
x^{(\alpha-1)/\alpha}z_*^{-2(1-\alpha)/(1+\alpha)}, 
\ee
(cf. eqn (3.16) without the term in curly brackets). If $\xi \rightarrow \infty$, then
$\dot{x}/x$, $\dot{S}/S$ and
$\xi$  must all diverge at the same rate (i.e. there is
no self-consistent solution with $\dot{x}/x << \dot{S}/S$ or $\dot{x}/x >> \dot{S}/S$). Hence,
\be
\frac{\dot{x}}{x} \sim x^{(\alpha-1)/\alpha},
\ee
which integrates to
\be
x = (d_1 + d_2 ln z)^{\alpha/(1-\alpha)}.
\ee
To ensure $\xi \rightarrow \infty$ as $z \rightarrow z_*$, one requires $x \rightarrow 0$ for $\alpha >0$ or 
$x \rightarrow \infty$ for $\alpha <0$. In both cases, one needs $d_1 =- d_2 ln z_*$ and hence
\be
x \sim [ln (z/z_*)]^{\alpha/(1-\alpha)}.
\ee
Defining
\be
L \equiv ln (z/z_*),
\ee
it is easy to show that the only consistent solution is then
\be
S = A L^{2/3(1-\alpha)},\;\;\;x = B L^{\alpha/(1-\alpha)}
\ee
where A and B are constants related by eqns (2.14, 2.15) and eqn (3.24). 
Thus the scale factor goes to zero, while the density
diverges for
$\alpha >0$ and goes to zero for $\alpha <0$. However, we also have 
\be
V\sim L^{-(1+3\alpha)/3(1-\alpha)}
\ee
and this diverges as $z\rightarrow z_*$ (as required) only for $\alpha >-1/3$. Eqn (2.14) implies 
\be 
M \sim L^{-2/3(1+\alpha)},
\ee
so $M \rightarrow \infty$ at $z=z_*$ and $MS \rightarrow$ constant. Eqn (2.14) also
implies that
$M$ is positive providing the sign of $\dot{S}/S$ is positive, so this requires that $z\rightarrow z_*$ from {\it above}. Note
that eqn (3.24) yields the same relation between A, B and $z_*$ as eqns (2.14, 2.15), so these solutions are described by two
independent parameters. Eqns (2.10) and (2.11) imply that the metric components are
\be
e^{\nu} \sim L^{\alpha/(1-\alpha)}\rightarrow 0,\;\;\
 e^{\lambda} \sim L^{-\alpha/3(1-\alpha)} \rightarrow \infty,\;\;\
R \sim L^{2/3(1-\alpha)}\rightarrow 0.
\ee
corresponding to a singularity of infinite density (cf. Schwarzschild). 

If $\xi \rightarrow 0$ or $\xi_*$ (constant), it can be shown that there is no consistent solution.

\vskip .3in
$\bullet$ $V\rightarrow V_*$ as $z \rightarrow z_*$

In this case, eqn (3.1) implies
\be
\frac{\dot{x}}{x} = \frac{2V^2\alpha}{V^2-\alpha}\frac{\dot{S}}{S} + \frac{2\alpha^2}{1+\alpha}\frac{1}{V_*^2-\alpha} -
\xi, \;\;\;
\xi \equiv \frac{\alpha (1+\alpha)\beta^2 V_*^2}{V_*^2-\alpha} z_*^{-2(1-\alpha)/(1+\alpha)} x^{(\alpha -1 )/\alpha}, 
\ee
where $V$ rather than $V_*$ appears in the first term on the right because the product $(V^2-V_*^2)(\dot{S}/S)$ may tend to
a finite limit. If $\xi
\rightarrow 0$, this implies
\be
 x \sim S^{2V_*^2\alpha/(V_*^2-\alpha)}
\ee
providing
$\dot{x}/x$ and $\dot{S}/S$ diverge. However, unless $\dot{V}/V \rightarrow \infty$, eqn (2.16)
also implies $x \sim S^{2\alpha/(1-\alpha)}$, so we require $V_*^2=1$. Eqn (2.12)
then becomes
\be
\ddot{S} + \left(\frac{1-4\alpha - \alpha^2}{1-\alpha^2}\right) \dot{S}- \frac{4\alpha}{1-\alpha}\frac {\dot{S}^2}{S} =0.
\ee
This is of the form of eqn (3.2), so the exact solution for $\alpha \ne 1/5$ and $\alpha \ne \sqrt{5}-2$
is given by
\be
S = [ c_1 + d (z/z_*)^{(\alpha^{2}+4\alpha-1)/(1-\alpha^{2})}]^{(1-\alpha)/(1-5\alpha)}.
\ee
In order to have $\xi \rightarrow 0$ as $z \rightarrow
z_*$, eqn (3.33) implies that one needs $S$ to go to infinity, whatever the sign of $\alpha$. One therefore requires
$d=-c_1$ and 
$\alpha >1/5$. This excludes $\alpha<0$ and also the special cases $\alpha =1/5$ and  $\alpha = \sqrt{5}-2$. Noting that to
first order
\be
1-\left(\frac{z}{z_*}\right)^{\frac{\alpha^{2}+4\alpha-1}{1-\alpha^{2}}}
\sim \left(\frac{\alpha^{2}+4\alpha-1}{1-\alpha^{2}}\right) L,
\ee
we obtain the asymptotic form
\be
S =A L^{(1-\alpha)/(1-5\alpha)},\;\;\;x = B L^{2\alpha/(1-5\alpha)}
\ee
where $A$ and $B$ are constants. Thus the scale factor diverges and the density goes to zero.
The condition $V_*=1$ gives a 
relationship between the constants $A$, $B$ and  $z_*$, so these solutions are
described by two parameters. 

Eqn (2.14) implies that 
\be
M \sim L^{(1-\alpha)/(5\alpha -1)},
\ee
so these solutions have zero mass at $z=z_*$ and $MS$ tends to a constant. Furthermore the sign
of $\dot{S}/S$ implies that $z$ must approach $z_*$ from {\it below} in order
that the mass be positive. Eqn (2.15) can be written as
\be
M = \frac{1}{2} + \frac{1}{2}\gamma^2 x^{-2/\alpha} S^6 \left\{\left(\frac{\dot{S}}{S}\right)^2(V^2-1) - 
\frac{2\dot{S}}{S} - 1\right\}.
\ee
Since $x^{-2/\alpha}S^6 \sim L^{2(3\alpha -1)/(5\alpha-1)}$ goes to infinity for $\alpha<1/3$ and zero for 
$\alpha >1/3$, one requires the term in curly brackets to go to to zero and infinity, respectively, in these two cases.
However, the last term in eqn (3.40) also scales as
\be
L^{(\alpha-1)/(5\alpha -1)} \left[\frac{\dot{S}}{S}(V^2-1) - 2 - \frac{S}{\dot{S}}\right],
\ee
so we need the term in square brackets to go to zero as $L^{(1-\alpha)/(5\alpha -1)}$ in both cases. We therefore need 
\be
\frac{\dot{S}}{S}(V^2-1) \rightarrow 2.
\ee
Using eqns (2.16) and (3.33), together with the relation
\be
\frac{V^2}{V^2-\alpha} \approx \frac{1}{1-\alpha}\left[1-\frac{\alpha}{1-\alpha}(V^2-1)\right],
\ee
one then obtains
\be
\frac{\dot{V}}{V} \rightarrow \frac{1-5\alpha}{1-\alpha}.
\ee
Condition (3.41) also
determines the second order terms in the expressions for
$x$ and $S$ but does not impose any further relationship
between $A$, $B$ and $z_*$.
Note that the metric can be written as
\be
ds^2 \sim L^{4\alpha/(1-5\alpha)} [dt^2 - dr^2 -r^2 L^{2(3\alpha -1)/(5\alpha-1)}d\Omega^2].
\ee
This resembles the open Friedmann model and is related to 
Minkowski spacetime by a time-transformation (as in the Milne model). These solutions can therefore be regarded 
as asymptotically flat, or, more precisely, as aymptotically Schwarzschild
since the mass ($\sim MS$) tends to a non-zero constant.

If $\xi \rightarrow \infty$ or constant, it can be shown that there is no consistent solution.

\vskip .3in
$\bullet$ $V\rightarrow V_*$ as $z \rightarrow \infty$ 

In this case, we require $\dot{V}\rightarrow 0$ and eqn (2.16) then implies 
\be
\frac{\dot{S}}{S} = \frac{1}{2}\left(\frac{1-\alpha}{\alpha}\right)\frac{\dot{x}}{x} + \frac{1}{2}\left(\frac{1-\alpha}
{1+\alpha}\right).
\ee
The exact solution of this differential equation is
\be
S^{2} = S_0 z^{\frac{1-\alpha}{1+\alpha}} x^{\frac{1-\alpha}{\alpha}}
\ee
which can then be combined with eqn (3.1) to give
\be
\frac{\dot{x}}{x} = \frac{V_*^2(1-\alpha)+2\alpha}{(V_*^2-1)(1+\alpha)}-\xi',\;\;\;
\xi' \equiv \frac{(1+\alpha)\beta^2 V_*^2}{V_*^2 -1} z^{-2(1-\alpha)/(1+\alpha)} x^{(\alpha -1 )/\alpha} 
\ee
where $\xi'$ differs slightly from $\xi$. If $\xi' \rightarrow 0$, this implies
\be
x = B z^m,\;\;\;m = \frac{V_*^2(1-\alpha)+2\alpha}{(V_*^2-1)(1+\alpha)}
\ee
and eqn (2.16) then gives
\be
S = A z^n,\;\;\;n \equiv \frac{(1-\alpha)(V_*^2+\alpha)}{2\alpha(1+\alpha)(V_*^2-1)}
\ee
where $A$ and $B$ are integration constants. For $\alpha>0$, these solutions are consistent with the condition $\xi'
\rightarrow 0$ as $z\rightarrow \infty$ providing $x \rightarrow \infty$ and, from eqn (3.48), this requires
$V_*^2 >1$. This in turn implies $S\rightarrow \infty$. Eqn (2.14) gives
\be
M \sim \frac{(V^2-\alpha)(1+\alpha)}{2\alpha(V^2-1)}z^{-[V_*^2(1-\alpha)+1+3\alpha]/(V_*^2-1)(1+\alpha)}
\ee
and so the mass is positive and tends to zero for $V_*^2>1$. On the other
hand, eqns (2.15), (3.48) and (3.49) imply
\be
M - \frac{1}{2} \sim  z^{[V_*^2(1-\alpha)-\alpha(1+3\alpha)]/
(V_*^2 -1) \alpha (1+\alpha)} \left\{\left(\frac{\dot{S}}{S}\right)^2(V^2-1) 
- \frac{2\dot{S}}{S} - 1\right\}.
\ee
If the exponent of $z$ in this expression is positive, $M$ can go 
to zero only if the term in curly brackets does and this requires
$\dot{S}/S \rightarrow 1/(V_*-1)$. Equating this with the value of $\dot{S}/S$ 
implied by eqn (3.48) gives a quadratic equation for $V_*$:
\be
 (1-\alpha) V_*^2 -2\alpha(1+\alpha)V_* - \alpha(1+3\alpha) =0
\ee
with the solution
\be
V_* = \frac{\alpha(1+\alpha) + \sqrt{\alpha(\alpha^3 - \alpha^2 + 3\alpha +1)}}
{1 - \alpha}.
\ee
As $\alpha$ decreases from $1$ to $0$, $V_*$ decreases from $\infty$ to $0$ and it
exceeds
$1$ (as required) only for
$\alpha>1/5$.  Note that eqn
(3.52) implies that the exponent of
$z$ in eqn (3.51) is indeed positive (as assumed).

The value of $V_*$ is also well-defined for $-1>\alpha>-1/3$. However, if one seeks solutions of
this kind, then the condition
$\xi'\rightarrow 0$ as
$z\rightarrow
\infty$  implies $x \rightarrow 0$ and eqn (3.48) would then give $1>V_*^2>-2\alpha/(1-\alpha)$. The lower limit is 
incompatible with eqn (3.53), so there are no $\alpha<0$ solutions. One can also formally obtain a negative
value of $V_*$ by taking the negative square root in eqn (3.53).
This is relevant if one is considering asymptotically KS solutions with $\alpha>0$.

Eqns (3.48), (3.49) and (3.53) impose a relationship between the constants $A$ and $B$ 
and so these solutions are described by one-parameter and they all tend to 
the same value of $V$ for a given $\alpha$. The requirement that the term in curly brackets in eqn (3.52) goes to
zero also determines the second order terms in the expansions for $x$ and $S$. 
These solutions, like the previous ones, are asymptotically flat: the metric 
components can be expressed as
\be
e^{\nu} \sim z^{V_*^2/(V_*^2 -1)},\;\;\;e^{\lambda} \sim z^{1/(V_*^2 -1)},\;\;\;R \sim z^{1/(V_* -1)}
\ee
and the metric can be reduced to the Minkowski form with a suitable change of 
coordinates.

If $\xi \rightarrow \infty$ or constant, there is no consistent solution.

\vskip .3in
$\bullet$ $V\rightarrow V_*$ as $z \rightarrow 0$

In this case, eqns (3.46) - (3.53) still apply but, for $\alpha>0$, the condition $\xi' \rightarrow 0$ requires $V_*^2<1$. 
Eqn (3.50) then implies that the mass is negative unless $V_*<1/\sqrt{\alpha}$. However,
eqn (3.53)
requires that one always has $V_*>1/\sqrt{\alpha}$, thus excluding this case. For $\alpha<0$, the condition $\xi' \rightarrow
0$ requires either $V_*^2>1$ or $V_*^2<-2\alpha/(1-\alpha)$. The first possibility is excluded because the mass is negative 
but,  
the second possibility is allowed providing $-1<\alpha<-1/3$, so that $V_*$ is defined.  

\setcounter{equation}{0}
\section{Discussion}

In this paper we have focussed on establishing the 
asymptotic forms of spherically symmetric perfect fluid similarity
solutions. The results are summarized in Table 1. Only those cases that give rise to self-consistent
and physically reasonable solutions have been described explicitly
(none of the remaining cases give rise to self-consistent solutions).
Although some of the details of the calculations have been omitted, the analysis presented can be regarded as a rigorous 
justification for the underlying assumption of our complete classification of positive $\alpha$ solutions in Carr \& Coley (1999b). 
Although our description of these solutions has been rather brief, their
physical properties have been further studied by Goliath et al. (1998a, 1998b) and
Carr et al. (1999). Our analysis has covered all equations of state 
with $|\alpha| < 1$ but the stiff fluid case
($\alpha = 1$) has not been explicitly included (although many of the results carry through
in this case); this is because different mathematical techniques are more appropriate in this case and the analysis
will be presented elsewhere.

It should be stressed that our analysis is not completely rigorous since we have implicitly assumed that certain
regularity conditions are satisfied.  For example, in Section 3 we always presupposed
that $V$ has a well-defined asymptotic limit.  Also our analysis has assumed certain properties for the derivatives
of small quantities. However, an asymptotic relationship of the form $X \gg Y$ does not necessarily imply
$\dot{X} \gg \dot{Y}$ asymptotically. In order to ``prove" the results of 
Section 3 we must show that this type of behaviour is not possible and this 
must be done on a case-by-case basis.
For example, suppose an asymptotic solution of the form $x \sim z^l$ is deduced. In principle 
\be
x \sim z^l(1+\varepsilon)
\ee
where
\be
\frac{\dot{x}}{x}= l +E,\;\;\; E \equiv 
\frac{\dot{\varepsilon}}{1+\varepsilon}.
\ee
However, the condition $\varepsilon \ll 1$
does not necessarily imply $E \ll 1$.  To prove the result, we 
must keep $\varepsilon$ and $\dot{\varepsilon}$ in all of the 
equations and show that the only 
self-consistent solutions must have 
$\varepsilon \ll 1$.
Suffice it to say that all cases have been checked and the 
only self-consistent solutions are those given in 
this paper. In fact, studying solutions of the form given by eqn (4.2) is precisely what we do in CC
in deriving solutions which are asymptotic to the Friedmann and Kantowski-Sachs models
(although in practice this calculation was divided up into
different separate cases than in that of section 3\---and 
this served as a further check of the validity of our results).

The results of Section 3, and the existence of well-defined 
limits and sufficient regularity, also follow from the mathematically
more rigorous dynamical systems analyses of Bogoyavlensky (1985)
and Goliath et al. (1998a, 1998b), at least in the
positive $\alpha$ case.  In these analyses,
monotone functions and Dulac functions were found to 
exist, thereby prohibiting periodic orbits and limit cycles
in the corresponding phase spaces.  This rules out the 
possibility of kinds of asymptotic behaviour different
from those discussed in this paper. 
Also, many of the results apply
to the case $\alpha = 0$ and are consistent with the results of the analytic
study of the dust models (Carr 1999); indeed, this 
serves as a further check of the validity of our results here. 

Finally, we should emphasis that our 
analysis has demonstrated that all self-similar spherically symmetric
perfect fluid solutions are asymptotically of power-law
form for $z \to \infty$ or $z \to 0$ and of log-power-law form
 for $z \to z_*$. This provides the basis of 
the work presented in our classification paper. 

\vskip .5in

{\large \bf Acknowledgments}

We thank Martin Goliath, Ulf Nilsson and Claes Uggla for helpful discussions. 
AAC was supported by the Natural Sciences and Engineering Research Council
of Canada.

\newpage

\noindent
{\large \bf References}

 \begin{enumerate}

\item[] G. V. Bicknell and R. N. Henriksen, 1978a, Ap. J. {\bf 219}, 1043.

\item[] O. I. Bogoyavlenski, 1977, Sov. Phys. JETP {\bf 46}, 634.

\item[] O. I. Bogoyavlenski, 1985, {\it Methods in the Qualitative Theory of  Dynamical Systems in Astrophysics and Gas Dynamics} (Springer-Verlag).

\item[] A. H. Cahill and M. E. Taub, 1971, Comm. Math. Phys. {\bf 21}, 1.

\item[] B. J. Carr, 1999, preprint.

\item[] B. J. Carr and A. A. Coley, 1999a, Class. Quant. Grav. {\bf 16}, R31.

\item[] B. J. Carr and A. A. Coley, 1999b, Phys. Rev. D, in press; gr-qc/9901050 (CC).

\item[] B. J. Carr and A. Koutras, 1993, Ap. J. {\bf 405}, 34.

\item[] B. J.  Carr and A. Yahil, 1990, Ap. J. {\bf 360}, 330.

\item[] B. J. Carr, A. A. Coley, M. Goliath, U.S. Nilsson and C. Uggla, 1999, gr-qc/9901031.

\item[] C. R. Evans and J. S. Coleman, 1994, Phys. Rev. Lett. {\bf 72}, 1782.

\item[] T. Foglizzo and R. N. Henriksen, 1992, Phys. Rev. D. {\bf 48}, 4645.

\item[] M. Goliath, U. S. Nilsson and C. Uggla, 1998a, Class. Quant. Grav. {\bf 15}, 167.

\item[] M. Goliath, U. S. Nilsson and C. Uggla, 1998b, Class. Quant. Grav. {\bf 15}, 2841.

\item[] C. W. Misner and H. S. Zapolsky, 1964, Phys. Rev. Lett. {\bf 12}, 635.

\item[] D. W. Neilsen and M. W. Choptuik, 1999, gr-qc/9812053.

\item[] K. A. Olive, 1990, Phys. Rep. {\bf 190}, 307.

\item[] A. Ori and T. Piran, 1990, Phys. Rev. D. {\bf 42}, 1068.

 \end{enumerate}


\end{document}